\documentclass[fleqn,12pt]{wlscirep}
\usepackage[utf8]{inputenc}
\usepackage[T1]{fontenc}
\usepackage[toc,page]{appendix}

\usepackage{authblk} 
\usepackage[numbers,sort&compress]{natbib}
\usepackage[parfill]{parskip}
\usepackage{hyperref}
\hypersetup{
  colorlinks,
  citecolor=.,
  linkcolor=.,
  urlcolor=.}

\renewcommand{\d}[2]{\frac{d #1}{d #2}} 
 
\definecolor{mypurple}{rgb}{0.82,0.02,0.48}
\definecolor{mygreen}{rgb}{0.1,0.6,0.2}
\definecolor{myblue}{rgb}{0.53,0.81,0.92}

\newcommand{\add}[1]{{#1}}

\usepackage{caption}
\title{Falling Through the Cracks: Modeling the Formation of Social Category Boundaries}

\author[1]{Vicky Chuqiao Yang}
\author[1]{Tamara van der Does}
\author[1]{Henrik Olsson}

\affil[1]{Santa Fe institute, 1399 Hyde Park Road, Santa Fe NM 87501, USA. }

\begin{abstract}
Social categorizations divide people into ``us'' and ``them,'' often along continuous attributes such as political ideology or skin color. This division results in both positive consequences, such as a sense of community, and negative ones, such as group conflict. Further, individuals in the middle of the spectrum can fall through the cracks of this categorization process and are seen as out-group by individuals on either side of the spectrum, becoming \textit{inbetweeners}. Here, we propose a quantitative, dynamical-system model that studies the joint influence of cognitive and social processes. We model where two social groups draw the boundaries between ``us'' and ``them'' on a continuous attribute. Our model predicts that both groups tend to draw a more restrictive boundary than the middle of the spectrum. As a result, each group sees the individuals in the middle of the attribute space as an out-group. We test this prediction using U.S. political survey data on how political independents are perceived by registered party members as well as existing experiments on the perception of racially ambiguous faces, and find support.

\end{abstract}

\begin{document}
\flushbottom
\maketitle

\vspace{0.1in} 

\section {Introduction}

Social categorization is a necessary and ubiquitous human social behavior, occurring on many attributes including race, gender, sexual orientation, and political ideology \cite{Lamont2002a}. On the one hand, social categorizing is essential for fulfilling a sense of community and a positive sense of self \cite{Tajfel1986}. On the other hand, \add{it} can fuel social conflicts by creating an ``us'' versus ``them'' mentality and \add{impacting} certain groups' access to economic and social resources \cite{bremmer2018us, Ashmore2004, Thoits1997, Roccas2008}. For example, the division between White and Black Americans has led to continuing discrimination and segregation long after the abolition of slavery \cite{Fox2012}. Recently, the divisions between Democrats and Republicans have created ``fear and loathing'' among U.S. voters \cite{iyengar2015fear}. 

A common theme in research on social categorization is to investigate the process of categorizing people as belonging to one's in-group or to an out-group. There is a vast literature on various intergroup biases such as in-group favoritism and out-group derogation \cite{brewer1999psychology, hewstone2002intergroup, dovidio2010intergroup}. Most of the research on social categories focuses on the end result of a social categorization process, where the typical assumption is that this process only leads to the perception of two groups \cite{dovidio2010intergroup,hewstone2002intergroup}. Moreover, most experimental work on individual classifications of race or gender attributes presents participants with pre-determined and forced choices \cite[for a review, see][]{bodenhausen2012social}. Similarly, in theories of social impression formation and social categorization, category representations and group motives are mostly treated as exogenous to the analysis with a fixed category structure in which individuals can be placed \cite{brewer1988dual, fiske1990continuum}. 

\add{As a result of the categorization process,} individuals can also ``fall through the cracks'' and not belong to any well-established social group. We refer to these individuals as \textit{inbetweeners}. Examples include mixed-race individuals who are considered neither truly Black nor White by members of either group, and political independents considered as ``other'' by both Democrats and Republicans. With demographic shifts, such as over ten million in the U.S. who identify with two or more races \cite{ACS2017} and the increasing gender non-binary population \cite{Schilt2017}, understanding how individuals fall through the cracks of categorization and the subsequent consequences are increasingly important. The existence of inbetweeners can be accommodated in existing models of social categorization by simply assuming that inbetweeners is a separate category. This assumption, however, does not address the question of how the boundaries of other categories are formed and how the inbetweeners category is created.  

Social categorization draws on both individual and social-level processes. A model for the formation of boundaries between categories must therefore include both levels. At the individual, or cognitive, level categorization decisions must relate to the distance between individuals. The influence of distance between individuals on categorization is part of many models of social categorization and social judgment \cite{galesic2018sampling, Smith1992}. At the social level, \add{this model} must take into consideration how other individuals influence the formation of boundaries between categories. The influence of others' beliefs and actions are well established in research on social categorization, social judgments, social learning, and belief formation \cite{cialdini1998social,festinger1954theory,ajzen1991theory}. Adopting categorization beliefs that are not supported by others in one’s immediate social environment can be costly, because it trigger\add{s} disapproval of others, withdrawal of cooperation, open conflict, or even ostracism \cite{williams2007ostracism,feinberg2014gossip}. In the social categorization literature, however, there are no quantitative models that integrate both the individual and social perspectives to predict how category boundaries are formed. 

In this paper we propose a quantitative, dynamical system model of social categorization that integrates cognitive and social processes. It predicts where category boundaries are placed and the occurrence of inbetweeners. Dynamical system models are useful for studying both formation and evolution processes because they enable tracking the feedback among many variables simultaneously. These models have been successful in explaining and predicting many complex social phenomena \cite{castellano2009statistical}, such as the extinction of minority languages \cite{abrams2003linguistics}, the decline of religious affiliation \cite{abrams2011dynamics}, the polarization in the U.S. Congress \cite{lu2019evolution}, and changes in party memberships in the UK \cite{jeffs2016activist}. In this model, the process of creating social category boundaries is influenced by individual-level cognitive processes and social processes. \add{In the cognitive process, individuals want to recognize whether others are similar to them.} This consideration is supported by several social-psychological mechanisms such as building successful collaborations \cite{Smaldino2012, Smaldino2018} as well as forming community and a positive sense of self \cite{Tajfel1986, Cerulo1997}. \add{However, remembering the exact distance between individuals is costly, therefore individuals will use categories as a summary of others' position.} At the social level, we assume that individuals consider the boundary choices of other group members. Those in the same group \add{want} to agree on who are \add{in} the in-group and who are \add{in} the out-group. After we present the mathematical model and its predictions, we \add{present some preliminary empirical validation} from the American National Election Studies (ANES) dataset and compare our results with findings from human behavior experiments.

\section{The Mathematical Model}

We model the formation of two social groups on a continuous attribute \cite{footnote1} and consider each group to have a boundary that divides the population into in-group and out-group. We will derive two governing equations, one for the boundary position of each group. We show in detail the derivation for one group, \add{which will be similar for the other group}. The derivation is achieved in two parts. In the first part (Section \ref{sec:indiErr}), we consider the individual-level cognitive process and derive the error in categorization for each individual. In the second part (Section \ref{sec:group_err}), we consider the group-level social process for agreeing with others in the same group, and derive the boundary position through optimizing the categorization error of the group. 

\begin{figure}[h!] 
    \centering
    \includegraphics[width = 0.75\columnwidth]{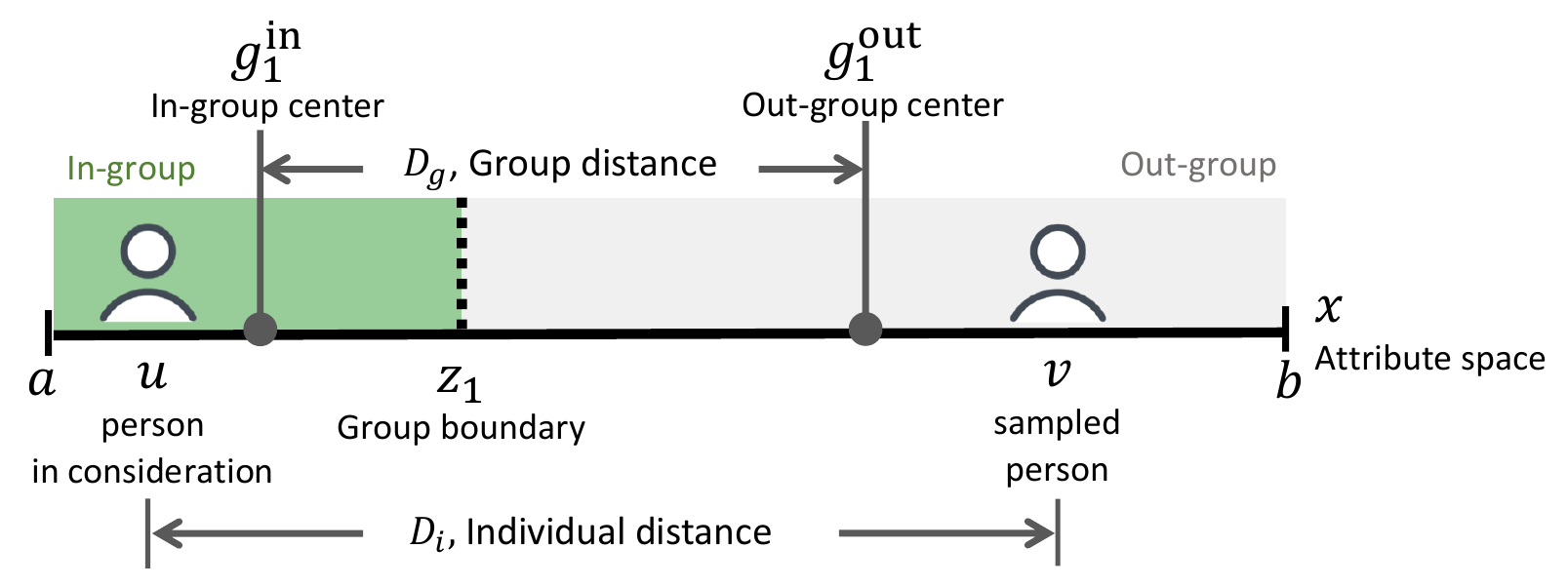}
    \caption{Illustration of the variables in the model. The illustration is presented from the perspective of a member  \add{of group 1}, interacting with an individual on the other side of the group boundary. The individual categorization error is the difference between individual distance and group distance. \add{Group 2} is not shown in this illustration.}
    \label{fig:cartoon}
\end{figure}

We denote the lower and upper bounds of the continuous attribute value $x$ to be $a$ and $b$, respectively, and the population distribution on the attribute to be $\rho(x)$. \add{We consider two groups forming on the continuous attribute space. One contains the left extreme of the attribute space, denoted as group 1, with boundary position $z_1$ that divides the in-group and out-group (see Fig.~\ref{fig:cartoon} for an illustration of the variables for this group). The other contains the right extreme of the attribute space, denoted as group 2, with boundary position $z_2$. The set-up for group 2 is symmetrical to that of group 1. We treat the boundary positions $z_1$ and $z_2$ as unknowns to be solved in the model.} 

\subsection{Individual-level cognitive process}\label{sec:indiErr}
The central insight from decades of research on categorization is that our cognitive system searches for patterns and structures \cite{rosch1975family}. The perception and cognitive representations of these patterns and structures can take many forms. In line with prototypical theories of category representations \cite{posner1968genesis,Rosch1973}, we assume a prototypical representation in the form of the mean position of a group. That is, we assume that all individuals categorized in the same group are perceived to have the group's mean position. For example, all individuals categorized under ``Democrat'' are perceived to have the mean position of all Democrats. Mathematically, the group positions for the in-group \add{($g_{\text{in}} (z_1)$)} and out-group \add{($g_{\text{out}}(z_1)$)} are defined as the the center of mass of the population distribution in each group,
\add{
\begin{equation}
g^{\text{in}}_1(z_1) = \frac{\int_a^{z_1} x \; \rho(x) dx}{\int_a^{z_1} \rho(x) dx} \;,\; \text{and}\;\; g^{\text{out}}_1(z_1) = \frac{\int_{z_1}^b x \; \rho(x) dx}{\int_{z_1}^b \rho(x) dx} \;. 
\end{equation}}



\add{We consider that individuals want to form categories of in-group and out-group because it is less costly than remembering individuals' precise attribute positions \cite{bodenhausen2012social}, and they want the categorization to reflect the actual differences in attributes as much as possible. Consider a person $U$ in the in-group \add{of group 1}, with position $u$ (\add{$u < z_1$}) on the attribute space. $U$ interacts with others on the attribute space through random sampling. Let $V$ be another individual on the attribute space (with position $v$) who interacts with $U$. $U$ observes the position of $V$ (for example, on the liberal to conservative scale).} $D_i$ denotes the distance between the two interacting individuals, $D_i = |u - v|$, and $D_g$ denote the distance between the two \add{individuals'} group positions (Fig.~\ref{fig:cartoon}). We define the categorization error for the interaction between \add{$U$ and $V$} to be the squared difference between the individual distance and the group distance, $(D_g - D_i)^2$. This error represents how much the group representation differs from the individual representation. The categorization error for $U$ perceiving \textit{all} sampled individuals is the integral of these errors with respect to $v$, weighted by the population density $\rho$,


%
\add{
\begin{equation} \label{eq:Indi_err0}
    err(u, z_1) =  \int_{a}^{b} (D_g - D_i)^2\rho(v) dv \;. 
\end{equation}
}

The term $D_i = |u - v|$ \add{is calculated} for all pairs of individuals. The term $D_g$ varies depending on if $v$ is \add{in the} in-group or out-group. If $v$ is \add{in the} in-group, both individuals are considered to be in the same group, $D_g = 0$. If $v$ is \add{in the} out-group, $D_g = |g_{\text{in}}- g_{\text{out}}|$. Combining with Eq.~\ref{eq:Indi_err0}, we have,  
\add{
\begin{equation} \label{eq:Indi_err1_1}
    err_1(u, z_1) =  \int_{a}^{z_1} |u - v|^2 \rho(v) dv + \int_{z_1}^{b} \left( |g^{\text{out}}_1 (z_1) - g^{\text{in}}_1(z_1)| - |u - v| \right)^2 \rho(v) dv \;. 
\end{equation}
}
 
The first term in Eq.~(\ref{eq:Indi_err1_1}) represents the error when the sampled person is \add{in the} in-group ($v<\add{z_1}$). The second term represents the case when the sampled person is \add{in the} out-group ($v>\add{z_1}$, the case illustrated in Fig.~\ref{fig:cartoon}). \add{The expression for the boundary of group 2, $z_2$ will be similar to the derivation process above. The domain of integration for the in-group will be changed from between $a$ and $z_1$ to between $z_2$ and $b$. The domain for the out-group will be changed from between $z_1$ and $b$ to between $a$ and $z_2$.}

Motivated by previous research \cite{Nosofsky1984, Shepard1987}, we considered an alternative formulation using similarity \add{instead of distance}. The model reaches the same main conclusion, though more mathematically involved, as shown in Appendix~\ref{app:similarity_model}.
%

\subsection{Group-level social process}\label{sec:group_err}

\add{In the social process, we consider individuals to be motivated to form a consistent category boundary with others in the same group. This, we assume, is a consequence of social learning and conformism where people are motivated by accuracy and affiliation goals \cite{Cialdini2004}. In our implementation, we approximate this process by assuming that individuals are concerned with agreeing with other in-group members \cite{soll2009strategies} and strive to minimize the average collective in-group categorization error. Individuals observe other members' categorizations and update their boundary position in the direction of the other members’ boundaries. Individuals do this while taking the individual categorization error (of Eq.~\ref{eq:Err1_1}) into account. This process repeats until the group arrives at one boundary position.} The average collective in-group error for group 1 is,
\add{
\begin{equation}\label{eq:Err1_1}
    Err_1(z_1) = \frac{1}{\int_a^{z_1} \rho(x)dx} \int_a^{z_1} err_1(u, z_1) \rho(u) du\;.
\end{equation}
}

Note that Eq.~(\ref{eq:Err1_1}) does not impose any preferences on group size. \add{Moreover, this formulation assumes that individuals weigh all other members of the in-group as equally important.} 
%
Finally, we consider \add{that} the group dynamically adjust\add{s} its boundary position to minimize the collective error, 
\add{
\begin{equation} \label{eq:dynamics}
\d {z_1} t = - k \d{Err_1(z_1)}{z_1}\;, 
\end{equation}
}
where $t$ is time, and $k$ is a constant that sets the time scale of the system. The intuitive understanding of Eq.~(\ref{eq:dynamics}) is that the category boundary evolves towards the direction that reduces the in-group's collective categorization error. \add{A similar process occurs for group 2, where the domain of integration in Eq.~\ref{eq:dynamics} is replaced by between $z_2$ and $b$, and $err_1(u, z_1)$ is replaced by $err_2(u, z_2)$.} \add{The social process above} can also be formulated as optimizations on the individual level, though with more complexity (Appendix~\ref{secApp:indModel}).


\section{Results}
\subsection{Model predictions}

\begin{figure}[htb!]
    \centering
    \includegraphics[width = 0.9\columnwidth]{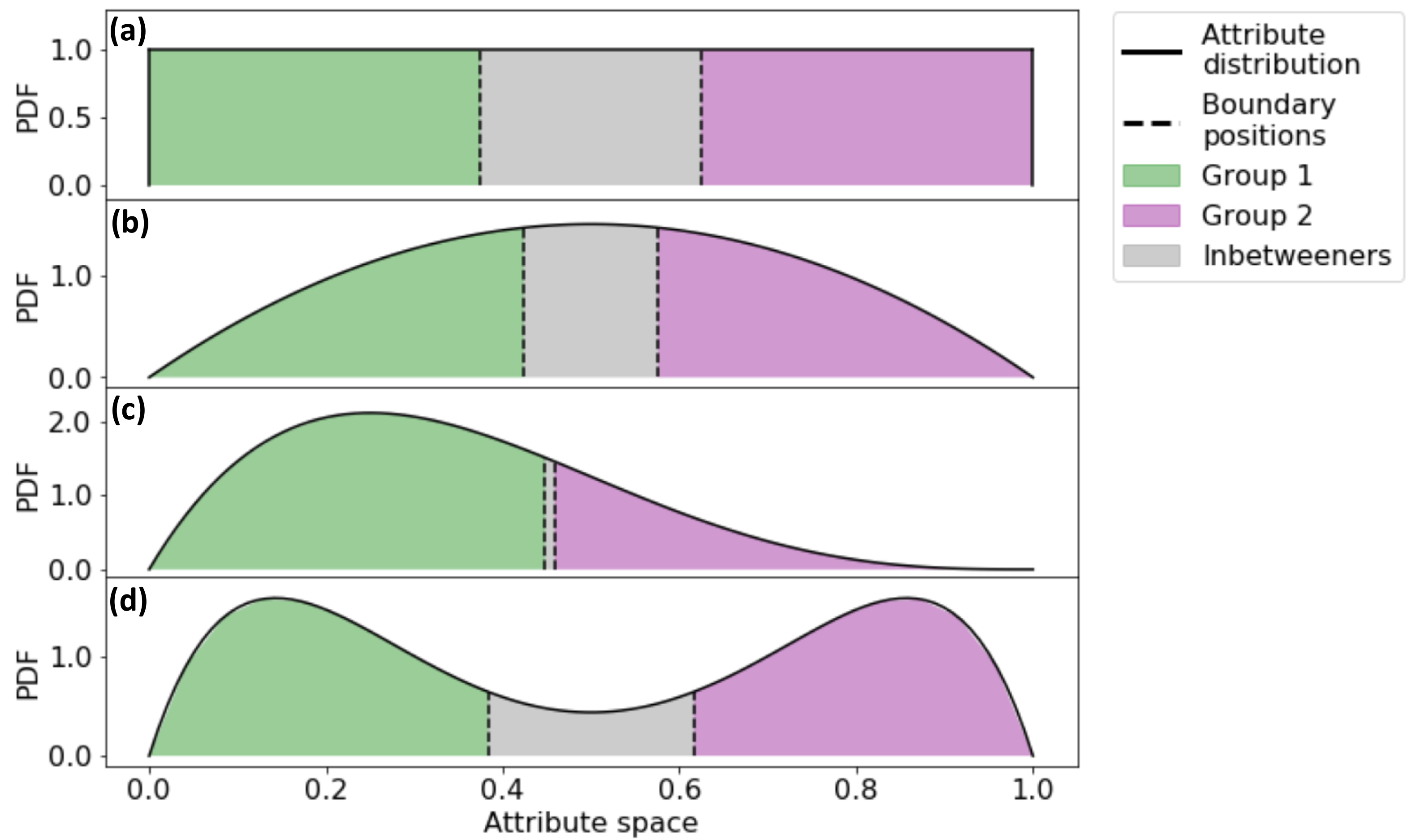}
    \caption{Stable fixed points of boundary positions for both groups: (a) for a uniform attribute distribution, (b) for the symmetrical Beta distribution with shape parameters $\alpha = 2$ and $\beta = 2$, (c) for the asymmetrical Beta distribution with $\alpha = 2$ and $\beta = 4$, and \add{(d) for a bi-modal distribution which is the sum of two Beta distributions, whose shape parameters are $\alpha = 2$, $\beta = 7$, and $\alpha = 7$, $\beta = 2$}. 
    }
    \label{fig:dist_model_results}
\end{figure}

We first present the results in the case where the attribute distribution $\rho(x)$ is a uniform distribution between $0$ and $1$ to demonstrate the behavior of the model. \add{With the uniform $\rho(x)$, the individual-level categorization error for members of group 1 is, 
\begin{equation}\label{eq:result_err1}
err_1(u, z_1) = u^2 - u z_1 + z_1^2/2 - z_1/4 + 1/12. 
\end{equation}
}
With this, we can analytically calculate the collective error, 
\add{
\begin{equation}\label{eq:Err_dist_Uni1}
    Err_1(z_1) = \frac{1}{3} z_1^2 - \frac{1}{4} z_1 + \frac{1}{12}\;.
\end{equation}
}

Equation (\ref{eq:dynamics}) has one stable fixed point, \add{$z^*_1 = 3/8 = 0.375$}, meaning the boundary position for group 1 stabilizes at 0.375: this group considers those with attribute value $x<0.375$ as in-group, and \add{those with attribute value} $x> 0.375$ as out-group. A same set of equations can be derived for group 2 (individuals on the right side of the spectrum). By symmetry, the preferred group boundary of group 2 is \add{$z_2^* = 0.625$}. This leads to individuals between 0.375 and 0.625 \add{to be} considered out-group by both social groups, which we refer to as inbetweeners (see Fig.~\ref{fig:dist_model_results}-(a)). 

The occurrence of inbetweeners is not unique to the uniform attribute distribution. \add{We now present} results obtained considering the attribution distribution $\rho(x)$ \add{as} a Beta distribution. The Beta distribution is parameterized by two positive shape parameters, $\alpha$ and $\beta$, with probability density function (PDF) \add{$f_{\text{beta}}(x, \alpha, \beta) = x^{\alpha-1}(1 - x)^{\beta - 1}/B(\alpha, \beta)$}, where $B(\alpha, \beta) = \Gamma(\alpha) \Gamma(\beta)/\Gamma(\alpha + \beta)$, and $\Gamma(\cdot)$ is the Gamma function. The distribution is defined for $x$ in the interval $[0, 1]$. We choose the Beta distribution because by adjusting the shape parameters we can produce a wide variety of unimodal distributions, both symmetrical and skewed. A number of real-world \add{attribute} distributions are known to be unimodal, such as political ideology of the U.S. public measured by positions on public policy issues \cite{fiorina2008political}. Panels (b) and (c) in Fig.~\ref{fig:dist_model_results} show two examples of the Beta distribution as attribute distribution $\rho(x)$, one symmetrical and one asymmetrical. In both cases, inbetweeners appear, though the location and size of the region vary with the distribution. \add{We have also analyzed the results for a bi-modal attribute distribution. We construct bi-modal distributions by summing two skewed Beta distributions that are symmetrical to each other. The PDF is $f_{\text{bimodal}}(x,\alpha ,\beta ) = 1/2 [f_{\text{beta}}(x,\alpha , \beta ) + f_{\text{beta}} (x,\beta ,\alpha )]$. Panel (d) in Fig.~\ref{fig:dist_model_results} shows the results for a bi-modal distribution with shape parameters $\alpha = 2$ and $\beta = 7$.}

\subsection{Validation with empirical findings}
\add{Our model predicts that those in the middle of the attribute space are seen as out-group by the two social groups as an outcome of the categorization process, becoming inbetweeners. While this paper's main focus is a contribution to theory, we look at data for preliminary validation of the model's predictions.}

We test our predictions using the American National Election Studies (ANES) dataset. \add{We focus on how Democrats and Republicans perceive those in the middle of the liberal-conservative attribute space. Since political independents tend to self-identify as being in the middle of the liberal-conservative spectrum, and Democrats and Republicans tend to select positions on either side (see Appendix~\ref{app:libConId}), we use political independents as an approximation for those in the middle of the attribute space.} We want to test if independents are perceived by both parties as part of the in-group (as favorably as \add{one's} own party), \add{as} the out-group (as unfavorably as the other party), or somewhere in between. If the perception \add{of} independents \add{is} similar to that of the other party, then the data supports our models' prediction. \add{We draw on research in social psychology \cite{fiske1982, fiske1990continuum} to argue that negative feelings are strongly driven by an out-group categorization. Even though feelings towards others are also driven by the difficulty in categorizing them \cite[for a review]{Lick2015}, we present here established categories thus removing any cognitive categorization work. We use feelings as an approximation for the categorization process and propose in the discussion other ways to test this model.} 


\begin{figure}[htb!]
    \centering
    \includegraphics[width = 0.6\columnwidth]{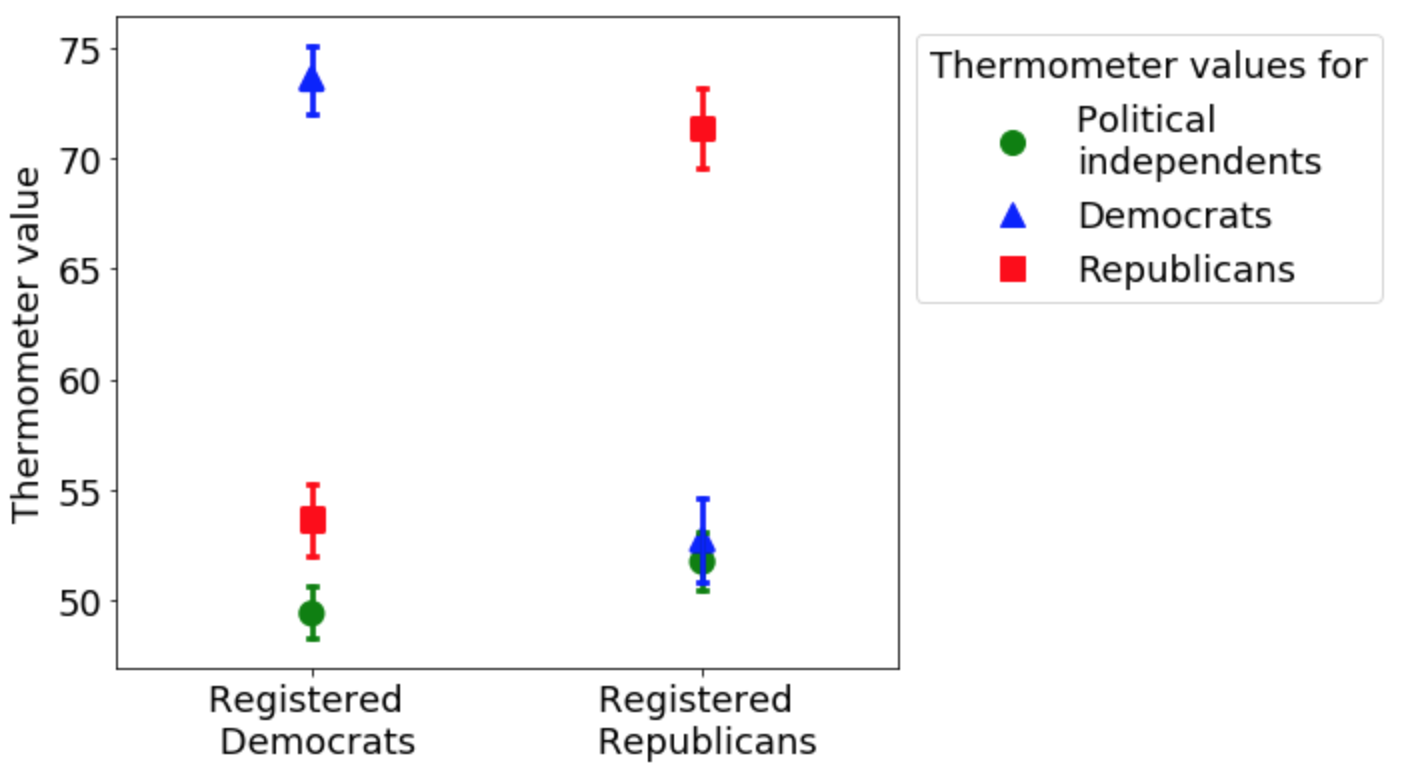}
    \caption{The mean thermometer values (reflecting feeling favorably or unfavorably) towards both political parties and political independents reported by registered party members (ANES). The error bars are 95\% confidence interval of the mean. For both registered Democrats and Republicans, political independents are perceived similarly compared to members of the other party, while members of own party are perceived more favorably.}
    \label{fig:ANESthermData}
\end{figure}
The ANES dataset is a nationally representative survey of political attitudes in the U.S. public. We use registered Democrat and Republican party members to represent the two groups on opposite sides of a continuous spectrum, as measured by self-reported party registration. We use a set of thermometer questions to measure attitudes towards Democrats, Republicans, and political independents.  In each thermometer question, participants are asked to report a number between 0 and 100---if they feel favorably about a group, a number greater than 50, and if they feel unfavorably about them, a number lower than 50 (see appendix \ref{app:ANES} for data source and questionnaire details). We use data from years 1980 and 1984, because the thermometer questions about political independents were only asked in these two years' surveys (N = 1,923). 

Figure~\ref{fig:ANESthermData} shows the mean thermometer values \add{towards} political independents, Democrats, and Republicans, reported by registered members of both parties. For both Democrats and Republicans, political independents are perceived similarly to members of the other party, while members of \add{one's} own party are perceived a lot more favorably. We perform a two-sided t-test and show that for Republican party members, the mean of thermometer values for Democrats and for political independents are indistinguishable ($p= 0.42$). The same test shows that for registered Democrats, the mean value for \add{feelings toward} Republicans is slightly higher than that of the political independents ($p < 0.001$). \add{One's} own party is perceived significantly more favorably than both the other party and the independents ($p < 0.001$). 

Beyond this ANES empirical test, our model's prediction is also in agreement with previous empirical studies on racial categorizations: in-group members tend to categorize ambiguous individuals as out-group, a process known as the in-group over-exclusion effect \cite{bodenhausen2012social, Leyens1992}. The following \add{four} types of experimental studies have confirmed this phenomenon. First, using established racial categories, perceivers in the U.S. tend to categorize racially ambiguous individuals as the out-group \cite{Gaither2016, Peery2008}, which was replicated in South Africa \cite{Pettigrew1958} and Italy \cite{Castano2002}. Second, using memory tests, an experimental study \cite{pauker2009} finds that racially ambiguous faces are perceived as out-groups by mono-racial individuals. Third, using open-ended categorization, a recent study points out that perceivers \add{tend to} use a third category (in this case, Hispanic or Middle Eastern) for racially-ambiguous individuals (who were mixed Black and White) \cite{Nicolas2019}. \add{Finally, studies measuring feelings towards bi-racial individuals find that they are on average rated more negatively through the process of categorization \cite{Halberstadt2014, Freeman2016}.} Taken together, these studies suggest that racial groups tend to draw boundaries that exclude individuals of mixed races, supporting our model's prediction of inbetweeners. 
 

\section{Discussion}

We propose a dynamical system model that integrates cognitive and social processes to arrive at social categorizations. The model predicts that social groups tend to draw boundaries that are more restrictive than the median of the attribute spectrum. As a  consequence, those in the middle of the attribute space are excluded by both social groups, becoming inbetweeners. Our theoretical finding is supported by empirical analysis of attitudes towards politically independent individuals by registered Democrats and Republicans, as well as by previous empirical findings surrounding the in-group over-exclusion effect in racial categorization \cite{bodenhausen2012social, Leyens1992}. The prediction of the existence of inbetweeners is unique to our model. Our model provides a rare theoretical result on how inbetweeners can arise \add{through the process of} social categorization. Although this work dominantly uses data on political ideology and racial categorization, our results can be extended more generally to individuals in the middle region of attribute spaces, such as those who are gender non-binary or in interdisciplinary scientific fields. 

Our model is intentionally parsimonious, aiming to capture key cognitive and social processes. We show here that a simple model can capture the main aspects of social category boundary formation. We do not attempt to model the influence of the myriad of motivational factors investigated in the social categorization literature, such as self-image maintenance \cite{fein1997prejudice} or motivated reasoning with stereotypes \cite{kundra1999motivated}. We also leave out \add{many complex cognitive and social processes which can influence social categorization, such as individuals' previous experiences and implicit biases towards members across the attribute spectrum \cite{bodenhausen2012social, Castano2002, Freeman2016} or culturally-based group hierarchies \cite{Wimmer2013, Castano2002}.} Social boundaries also evolve over time. For example, the shifting demographics in the U.S. since the 1960s have extended racial category labels beyond the dichotomy of Black and White \cite{Bonilla-Silva2004}. The dynamical systems framework we propose can be used, in future research, to explore how demographic and cultural shifts lead to changes in social category boundaries. \add{In this manuscript, we study the rise of inbetweeners as a result of categorization. It is possible for this result to have further downstream consequences, such as forming a new group with other inbetweeners. It would be useful for future research to study how inbetweeners form new groups}. 

Our empirical validation shows initial support for the presented model, however the thermometer measures can only approximate how individuals categorize each other as in- or out-group. \add{Affect (feelings toward others) is the result of many categorization processes beyond boundary formation, such as the difficulty of categorizing the other \cite{fiske1990continuum, Lick2015}. Future research can benefit from investigating the perception of inbetweeners in an experimental setting, where the distribution of individuals in the attribute space is known and the outcome variable focuses on actual categorization and not affect. This could be achieved by asking individuals across the political spectrum to mark others as in- or out-group based on their policy views.}
 

Much significance of social categories is not in the categories themselves, but in how these categories affect how individuals are perceived and treated. Our model's prediction that individuals with characteristics in the middle of the attribute space ``fall through the cracks'' may affect many social processes. One speculative example is the disconnect between issue polarization and social polarization in the U.S. public. Previous empirical research has found that identification with political parties and antipathy towards the opposing party increased disproportionally compared to opinion on issues \cite{mason2015disrespectfully, fiorina2008_a, hetherington2009, hill2015}. Our model provides a possible explanation that political independents are perceived as out-group by both political parties. Motivated by the need for belonging and community, individuals holding moderate positions might decide to instead identify with one of the two polarized parties, despite misalignment on issue positions. Empirically testing how political independents are perceived and how this relates to social polarization can be an important direction for future research. 

\vspace{15pt}

\textbf{Acknowledgments.} We thank the 2018 Fall JSMF-SFI Postdoctoral Conference for allowing a research jam on the topic of categorical perception, where this project started. We thank Mirta Galesic and Sidney Redner for helpful feedback on the manuscript.

\appendix
\section*{Appendices}
	\setcounter{figure}{0}
	\setcounter{equation}{0}
	\renewcommand\thefigure{A\arabic{figure}}  
	\renewcommand\thetable{A\arabic{table}}  
	\renewcommand\theequation{A\arabic{equation}}

\section{Similarity-based model}\label{app:similarity_model}
In the main text, we considered a model where differences between individuals are measured by the one-norm distance. This choice is driven by deriving a parsimonious model that captures the key behavior of the system \add{and it is a common assumption in models that describe belief change from social influence (for example, in models based on variants of DeGroot learning \cite{DeGroot1974}). However, research on multidimensional scaling and categorization suggests that a plausible model of how distance relates to categorization involves \textit{similarity} \cite{Shepard1987}. Similarity describes how similar two attribute values are to the cognitive system, and is a nonlinear function of the distance. Here, we show that show our results of inbetweeners is not unique to the one-norm distance measure, but also arise with a measure of similarity. }

Empirically, similarity between two positions $x_1$ and $x_2$ (denoted as $s(x_1, x_2)$) can be described as a nonlinear decreasing function with the distance between them in the form of 
\begin{equation}
s(x_1, x_2) = \exp(- c |x_1 - x_2|)\;, 
\end{equation}
where $c$ is the sensitivity parameter---larger $c$ means individuals are more sensitive to differences in attribute space. Variants of this functional form are used in many categorization and social judgment models \add{ \cite[e.g.,][]{Nosofsky1984, Smith1992}.}

Consider the group containing the left extreme of the attribute space. Similar to the derivation in the main  text, the categorization error of an individual at position $u$, with group boundary \add{$z_1$} is
\add{
\begin{eqnarray}\label{eqApp:Indi_err1}
    err_1(u, z_1) &=& \int_{a}^{z_1}\left[s(g_{\text{in}}(z_1), g_{\text{in}}(z_1)) -  s(u, v)\right]^2 \rho(v) dv \\\nonumber
    && + \int_{z_1}^{b} \left[s(g_{\text{in}}(z_1), g_{\text{out}}(z_1)) - s(u,v)  \right]^2 \rho(v) dv  \;.
\end{eqnarray}}
The other equations of the model remain unchanged. 

\begin{figure}[htb!]
    \centering
    \includegraphics[width = 0.7\columnwidth]{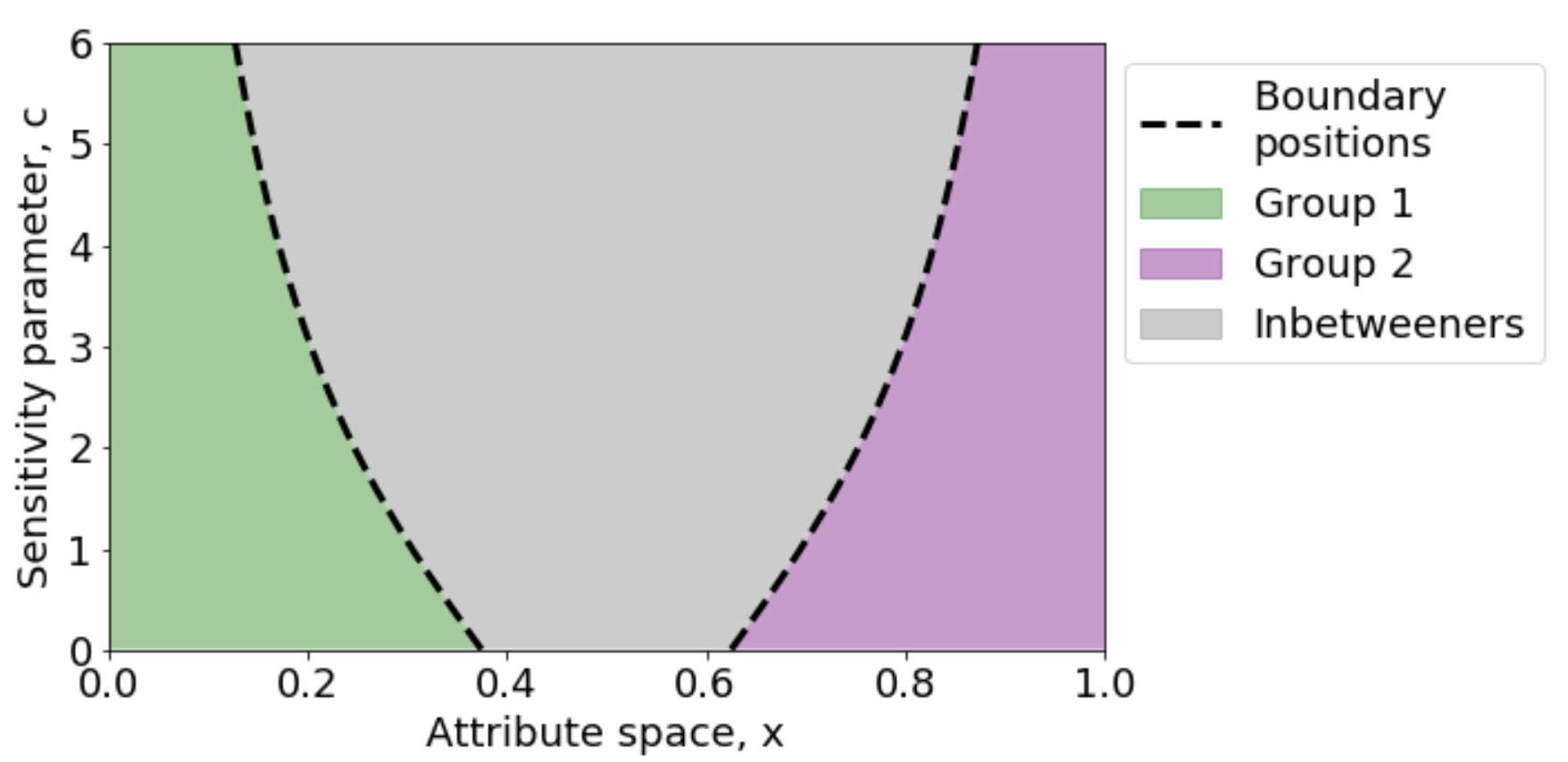}
    \caption{Stable fixed points of boundary positions for both groups, as a function of the sensitivity parameter, $c$. The dark grey region represents the inbetweeners. This result is for uniform attribute distribution $\rho(x)$ defined for $x$ between $0$ and $1$. }
    \label{fig:xstar_vs_c_uniform}
\end{figure}

The model's results for uniform distribution is shown in Fig.~\ref{fig:xstar_vs_c_uniform}, displaying solutions of boundary positions of the two groups (\add{$z_1^*$ and $z^*_2 = 1-z_1^*$}) as a function of $c$ . As $c$ increases, meaning people more sensitive to differences in the attribute space, the social group boundaries become more exclusive. For all $c >0$, the fixed point satisfies $z^*< 0.5$, meaning inbetweeners occur for all values of $c$. 
\section{Individual-level model} \label{secApp:indModel}
\begin{figure}[htb!]
    \centering
    \includegraphics[width = 1\columnwidth]{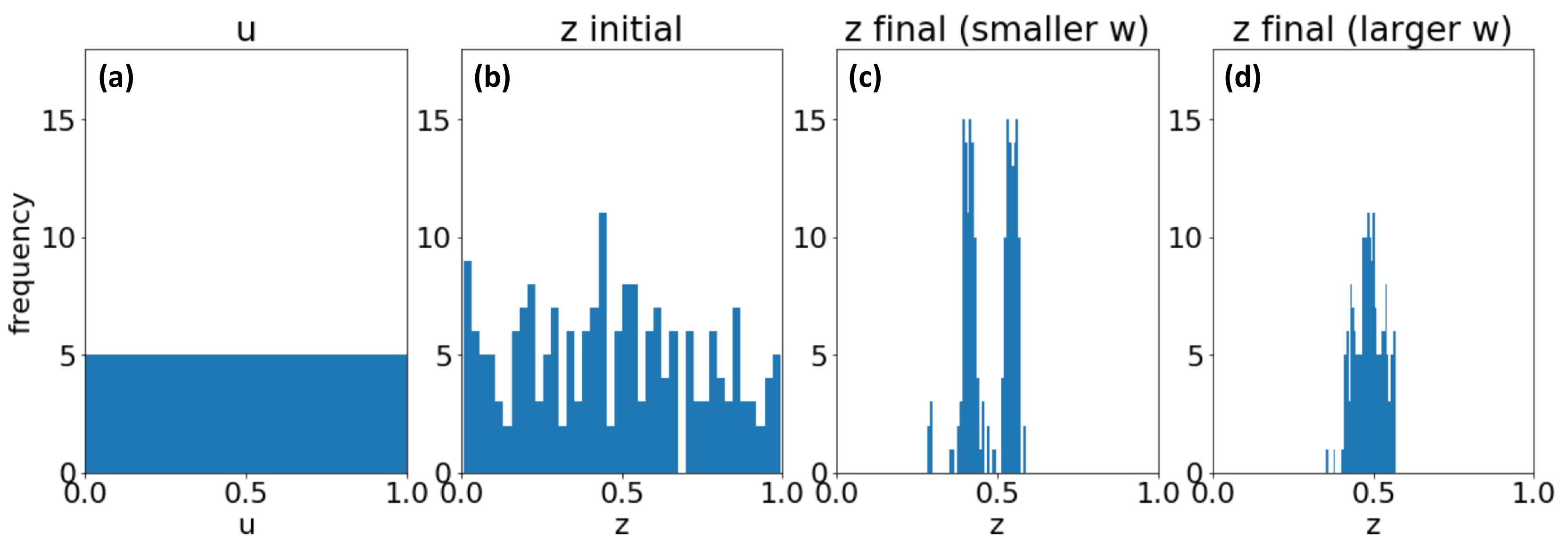}
    \caption{\add{Simulations of the individual process level model for 200 individuals. (a) The population distribution in the simulation, an evenly spaced population. (b) The initial condition of boundary condition, drawn from a uniform distribution. (c) distribution of the individuals' boundaries after 500 time steps for $w = 0.2$. The distribution of the boundaries is bi-modal. (d) The distribution of the individuals' boundaries after 500 time steps for $w = 0.3$. The distribution is uni-modal.}}
    \label{figApp:indModelResult}
\end{figure}

In the main text, we used a group-level model in Section~\ref{sec:group_err}, where the group optimizes a collective in-group error. Here, we show an alternative formulation, where we prescribe the social process explicitly on the individual level.

In the group-level formulation, we consider all in-group individuals to form the same category boundary. In the individual formulation, we consider each individual (with position $u$) having a boundary position ($z(u)$), which can be distinct from that of other individuals. The part of the attribute space on the same side of the boundary is considered in-group for the individual, and the other side out-group. \add{Note that here we no longer predict two group boundaries in attribute space like in the main text, but a distribution of boundary positions.} 

Each individual's utility consists of two components. One is the categorization accuracy, similar to Eq.~\ref{eq:Indi_err1_1}, and the other is agreement with the boundary position of others in the same group. The categorization error for the individual at position $u$ with boundary position $z$ is, 
\begin{equation}\label{eqApp:indErr}
    err(u, z) = \sum_{v \in \text{ingroup}}(u- v)^2 + \sum_{v \notin \text{ingroup}} \left(|g_{\text{in}}(z) - g_{\text{out}}(z)| - |u, v| \right)^2
\end{equation}
Equation~\ref{eqApp:indErr} \add{computes the individual categorization error the same way as} Eq.~\ref{eq:Indi_err1_1} in the main text, but written in a discrete formulation, \add{where the condition for $v$ being in group is: (1) if $ u>z$, $v>z$; and (2) if $u < z$, $v< z$. Otherwise, $v$ is outgroup. }


The cost (negative utility) function for each individual is,  
\add{
\begin{equation} 
f(u, z) = \frac{w}{N} \; err(u, z) + \frac{1-w}{N_{in}} \sum_{v \in \text{ingroup}}(z(u) - z(v))^2\;.
\end{equation}
}
The first term is the error from categorization accuracy, and the second term is the cost from disagreeing with others in the in-group. The normalization term $N$ is the total number of individuals in the system, and $N_{in}$ is the number of people in the in-group. Parameter $w$, which is between $0$ and $1$, is the weight of the categorization accuracy compared to the need for having a boundary similar to others of the in-group. The dynamics for each individual is
\begin{equation}
    \frac{d z}{d t}= -\frac{\partial f(u, z)}{\partial z}\;.
\end{equation}
This model generates a distribution of boundary positions. We simulate the system in discrete time, and an example is shown in Fig.~\ref{figApp:indModelResult}. With a uniform distribution in attribute space, for parameters $w$ smaller than a threshold, the system evolves to have a bi-modal distributed boundary positions. \add{In numerical simulations, the transition between uni-modal to bi-modal occurs between $w = 0.2$ and $0.35$, while the precise critical $w$ value fluctuates within this range with the initial condition of the simulation. Intuitively, the presence of a transition can be understood as the following. When $w = 1$, where individuals are not affected by others at all, each individual's boundary is at their preferred position, and the distribution has no bi-modal behavior. When $w = 0$, individuals care only about social agreement, and two peaks would form on the attribute space. Thus a transition between these two kinds of distributions occurs for some $w$ between 0 and 1, and the bi-modal boundary distributions occur when agreeing with others becomes more important, which would be similar to the model in the main text (group's members must agree on one position).}

\add{We can estimate the boundaries of the two groups using the peaks of the distributions. In the bi-modal distribution (smaller $w$, suggesting social influence is more important) , we approximate the two group boundaries with the two peaks of the distribution, and the individuals between these two peaks become inbetweeners, consistent with our findings in the main text. However, when individuals are less motivated by social influence (larger $w$), the distribution becomes unimodal. With estimating group boundaries with the peaks of distribution, the two group boundaries collide, and we no longer have inbetweeners. }

\section{American National Election Studies Data} \label{app:ANES}
The American National Election Studies data used in this paper is the cumulative data file of 1940-2016, May 31, 2018 version. It was downloaded from \url{https://electionstudies.org/project/anes-time-series-cumulative-data-file/} on Oct 4, 2018. 

The phrasing of the thermometer questions is as follows: ``We'd also like to get your feelings about some groups in American society. When I read the name of a group, we'd like you to rate it with what we call a feeling thermometer. Ratings between 50 and 100 degrees mean that you feel favorably and warm toward the group; ratings between 0 and 50 degrees mean that you don't feel favorably towards the group and that you don't care too much for that group. If you don't feel particularly warm or cold toward a group you would rate them at 50 degrees. If we come to a group you don't know much about, just tell me and we'll move on to the next one. Using the thermometer, how would you rate the following...?'' The three groups used in our analysis are ``Democrats,'' ``Republicans,'' and ``political independents.''

\section{\add{American National Election Studies Data}} \label{app:libConId}
\begin{figure}
    \centering
    \includegraphics[width = 0.5\textwidth]{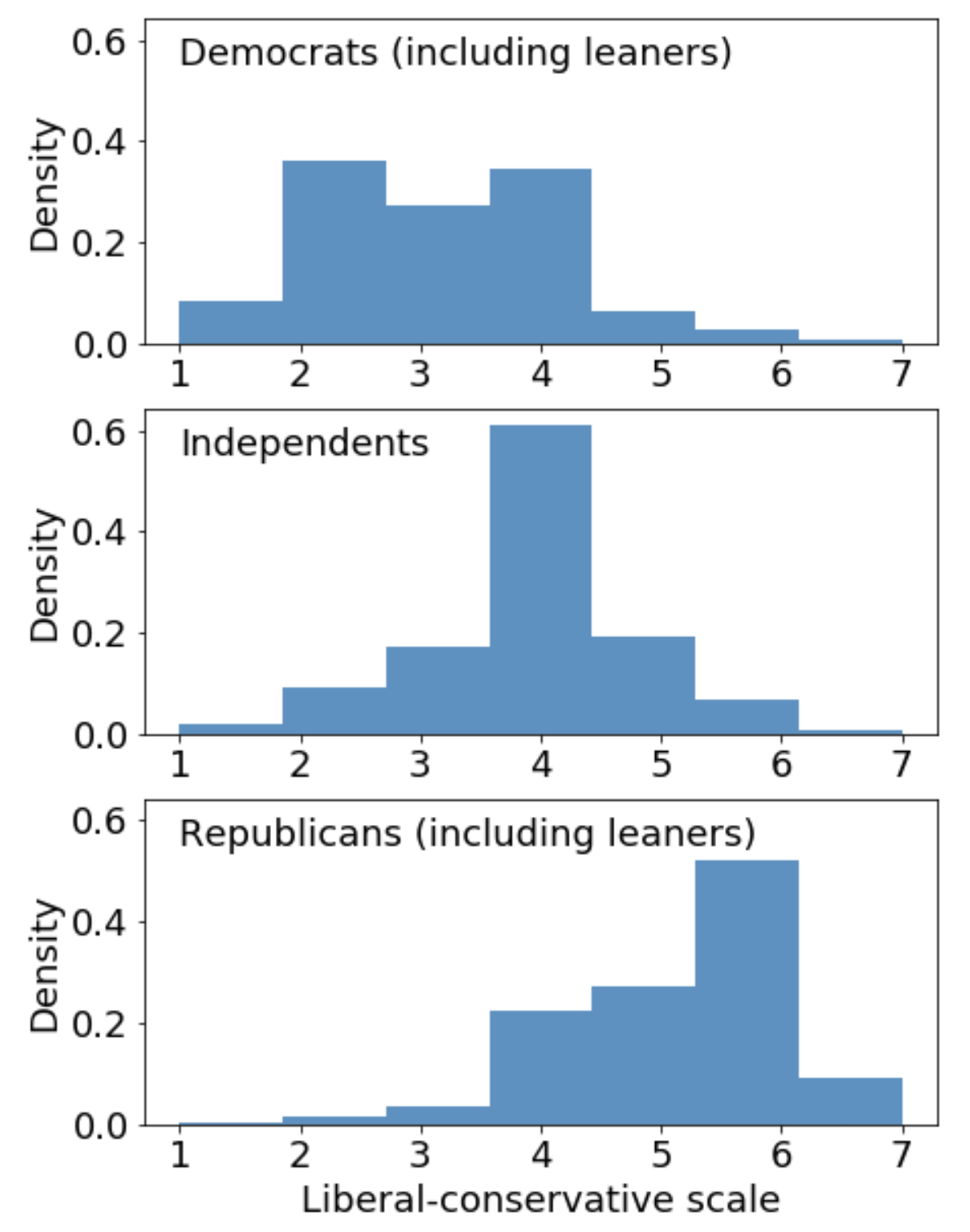}
    \caption{\add{Self-identification on the liberal-conservative attribute space for Democrats, including leaners (top panel), Political Independents (middle panel), and Republicans, including leaners (bottom panel). Scale value 1 = extremely liberal, and 7 = extremely conservative.}}
    \label{fig:app_libConId}
\end{figure}

\add{In the main text, we use political independents to approximate those who are in the middle of the liberal-conservative attribute space. In Fig.~\ref{fig:app_libConId}, we show normalized histograms of self-reported liberal-conservative identification Democrats, Republicans and independents from ANES data. Indeed, independents tend to identify as the middle of attribute space, compared by Democrats and Republicans. 
}

\begin{thebibliography}{10}
\bibitem{footnote1}
	\bibinfo{note}{For simplicity, we present the model for two groups, which
		already leads to complex model behavior. The modeling framework is readily
		extendable to $n$ groups.}
\end{thebibliography}
\end{document}